\documentclass[english]{article}
\usepackage[T1]{fontenc}
\usepackage[latin9]{inputenc}
\usepackage{graphicx}
\usepackage{esint}

\makeatletter
\newcommand{\lyxaddress}[1]{
\par {\raggedright #1
\vspace{1.4em}
\noindent\par}
}
\newenvironment{lyxlist}[1]
{\begin{list}{}
{\settowidth{\labelwidth}{#1}
 \setlength{\leftmargin}{\labelwidth}
 \addtolength{\leftmargin}{\labelsep}
 }}
{\end{list}}

\makeatother

\usepackage{babel}
\begin{document}

\title{Clustering in complex ionic liquids in two dimensions}

\author{Aurélien Perera$^{1}$and Tomaz Urbic$^{2}$}
\maketitle

\lyxaddress{$^{1}$Laboratoire de Physique Théorique de la Matière Condensée
(UMR CNRS 7600), Université Pierre et Marie Curie, 4 Place Jussieu,
F75252, Paris cedex 05, France.}

\lyxaddress{$^{2}$Faculty of Chemistry and Chemical Technology, University of
Ljubljana, Vecna pot 113, 1000 Ljubljana, Slovenia.}
\begin{abstract}
Two-dimensional ionic liquids with single site anion and cation-neutral
dimer are studied by computer simulations and integral equation techniques,
with the aim of characterizing differences with single site anion-cation
mixtures, and also with three dimensional equivalents of both models,
in order to see the competition between the Coulomb interactions and
the clustering restrictions due to reduced dimension. We find that
the addition of the neutral site to the cation suppresses the liquid-gas
transition which occurs in the case of the monomeric Coulomb system.
Instead, bilayer membrane type ordering is found at low temperatures.
The agreement between the structural correlations predicted by theory
and the simulation is excellent until very close to the no-solution
region predicted by the theory. These findings suggest various relations
between the nature of the clustering at low temperatures, and the
inability of the theory to enter this region.
\end{abstract}

\section{Introduction}

Coulomb interactions in three dimensions (3D) and two dimensions (2D)
differ in an interesting manner, not only because the former has a
$1/r$ form, while the latter has a $\ln(r)$ form\cite{10Lubensky},
but because the first diminishes with inter particle distance, while
the second increases. This second behaviour is contrary to the intuition.
As a consequence, the 2D Coulomb interaction has opposite sign than
in the 3D case\cite{12Hansen}. Indeed, for charges of opposite sign
to attract, the 2D Coulomb interaction must be endowed with a positive
sign, which means that particle come closer to each other because
they repel even more at large distances. Similarly, like charges repel
each other at short distances because in fact they attract each other
if they are far spaced. Despite this fundamental difference, we have
recently shown\cite{13our2Dionic} that both systems have similar
structure in the fluid phase, namely the charge ordering property\cite{14early}. 

Charge order is principally a short range feature, which enforces
a checker-board appearance of the local order\cite{16myPCCPchord},
due to optimizing of attraction/repulsion of the pair interactions.
It leads to square lattice crystal order in 2D in the high density
low temperature regime. Some of this order is locally retained in
the liquid phase at higher temperatures. Even in the low density regime,
where clustering dominates the structure of the system, the clusters
obey charge order. In the same recent paper\cite{13our2Dionic}, we
have shown that, replacing the 2D Coulomb interaction by the 3D screened
Coulomb form, retained a strikingly similar local charge order, both
in the liquid and gas phases. This finding enforces the idea that
local order is mainly ruled by the strong pair interaction, despite
the very different forms and signs of the full respective Coulomb
interactions. 

It is with this idea in mind that we wish to study how charge ordering
in 2D is affected by the presence of neutral sites attached to the
cation. This is motivated by the so-called room temperature ionic
liquids (RTIL), such as ethylammonium nitrate, for example, which
are liquid at room temperature\cite{18RTIL} when ionic system such
as NaCl are crystalline. We have previously argued\cite{20ourIL}
that it is the presence of inert sites attached to one of the charged
atoms, which allow to reach the liquid state at low temperature, by
hindering charge order to induce a crystalline state. In particular,
this hindered charge order produces local clustering of the charged
segments, which, in turn, produces a scattering pre-peak in the structure
factors. These effects are well documented for realistic RTIL in 3D
\cite{22pptriolo1,23pptriolo2,24ppcastner,25ppcastner2,26CCvoth1,27CCvoth2,28ribeiro}.
For the 2D case, it would be interesting to know how this screening
of the charge order, induced by the presence of neutral sites, is
affected, when we expect lesser possibilities of molecular conformations?
To this effect, we study here the influence of thermal disorder at
various temperatures, but also at various densities, in order to see
the influence of clustering. In addition, there are 2 other interesting
issues. The first issue concerns the existence of a liquid-gas coexistence
at low temperatures, which has been intensely studied for case of
the simple monomer system the so-called restricted primitive model
(RPM)\cite{291hansbaus,292levinFisher,293caillol,294pana,295juan,296Wolffram}
and its 2D version\cite{2990fisher,2991pana,2992marjolein,2993caillol,2994samaj,2995hansen,2996mak,2997lomba,2997tomaz,2998tomaz,2999tomaz}.
The second issue concerns the existence of a Kosterlitz-Thouless (KT)
transition\cite{300KT} in the very low density region, where no free
charges exist below the KT horizontal line in the (density, temperature)
phase diagram\cite{2995hansen,2997lomba,2997tomaz}. We examine here
how these 2 properties are affected by the presence of neutral sites. 

In the present study, we principally focus on the model illustrated
in Fig.1, namely a single site anion and linear molecular cation with
sites tangently attached to each other, with the cation site at on
end. Our study clearly indicates that 3D to 2D dimensional reduction
hinders the stability of the liquid phase at lower temperatures, such
that the molecular ionic system has a stable liquid state for temperatures
higher than the simple monomer ionic liquid. This is exactly the opposite
behaviour than in 3D. In other words, molecular complexification in
2D restricts the range of the liquid phase. Interestingly, layer-like
clustering is favoured at low densities, enforcing an a horizontal
asymptote in the (density, temperature) phase diagram for the stability
of the disordered phase with respect to the cluster ``phase''. Since
this cluster ``phase'' needs to be specified in relation to the
behaviour of the integral equation theory, we can only conjecture
about the underlying KT type behaviour in relation to the layer-like
association which occurs in this part of the phase diagram. 

\section{Models and technical details}

In a previous paper\cite{13our2Dionic}, we have considered charged
system of monomers of equal size, with the 3D form of screened Coulomb
interaction. In the present paper, as illustrated in Fig.1, we consider
a monomeric anion together with a dimer made of tangent spheres, one
being a cation and the other neutral. All sites are taken to be spheres
of diameter $\sigma$. The total site-site interaction reads
\begin{equation}
\beta v_{ij}(r)=Z_{i}Z_{j}\frac{T_{C}}{T}\frac{\exp(-r/\lambda)}{r/\sigma}+\frac{4T_{0}}{T}\left(\frac{\sigma}{r}\right)^{12}\label{int}
\end{equation}
where $\beta=1/k_{B}T$ is the Boltzmann factor, with T the temperature
expressed in Kelvin, $T_{C}=55700$K corresponds to the temperature
in the 3D Coulomb interaction \cite{20ourIL} for the choice $\sigma=3$\AA,
$T_{0}=100$K is arbitrarily chosen. The screening parameter is chosen
to be $\lambda=2$. All these parameters are the same as in Ref.\cite{13our2Dionic}.
The valences are $Z_{1}=-1$, $Z_{2}=+1$ and $Z_{3}=0$.

It is important to note that it would be incorrect to conclude that,
since we use screened Coulomb interactions, the charge influence is
minor. Indeed, in Ref.\cite{13our2Dionic}, we demonstrated that the
structure of the system, as witnessed by the correlation functions,
both in real and reciprocal space, show the same characteristics as
when unscreened Coulomb is used. This remark is even more important
since we are comparing the log-Coulomb of the 2D case with a screened
version of the 3D case. The reason for this strong influence of the
charges comes from the fact that the Coulomb interactions dominate
the short range ordering through the factor $T_{C}\gg T_{0}$ . Interestingly,
this effect is not only important in dense fluid, but also in the
gas phase, as we show below in the Results section.

\subsection{Monte Carlo simulations}

All Monte Carlo (MC) simulations are done in the canonical (NVT) ensemble
following the same protocol previously outlined in Ref.\cite{13our2Dionic}.
With one MC cycle consisting of a tentative move of N particles, $10^{6}$
equilibration moves and $10^{7}$ moves for statistics are performed
for each system. Cut off of the potential was half-length of the simulation
box. All simulations were performed with $N=100$ or $N=200$ molecules.
Increasing the number of particles had no significant effect on the
calculated quantities. The fact that smooth correlation functions
with very low noise level are obtained is strong indication that the
simulations are well converged.

\subsection{Integral equation theory}

Concerning the integral equation theory (IET) approach, we have used
the 2D site-site Ornstein-Zernike (SSOZ) formalism\cite{34chandler,35hansmac}
together with the hypernetted chain (HNC) closure\cite{35hansmac}.
The choice for this particular closure in place of others, such as
the mean spherical approximation (MSA), self-consistent closures\cite{2995hansen,2997lomba},
or the Hirata-Kovalenko type closure\cite{34KH}, is that the HNC
closure represents the first level of approximation where all correlations
higher than rank 2 are neglected\cite{35hansmac,36bfmt}. 

The SSOZ equation consists in the following matrix equation:

\begin{equation}
SM=I\label{ssoz}
\end{equation}
where the total structure factor matrix $S$ is given by

\begin{equation}
S=W+\frac{\rho}{2}H\label{sk}
\end{equation}
and

\begin{equation}
M=W^{-1}-\frac{\rho}{2}C\label{mk}
\end{equation}
The intramolecular part of the total structure factor is defined through
the matrix $W$ as

\begin{equation}
W=\left(\begin{array}{ccc}
1 & 0 & 0\\
0 & 1 & J_{0}(k\sigma)\\
0 & J_{0}(k\sigma) & 1
\end{array}\right)\label{wk}
\end{equation}
where $J_{0}(x)$ is the zeroth-order integer Bessel function. $\rho=N/V$
is the total density of the system. The matrices $H$ and $C$ with
respective elements $\tilde{h}_{ij}(k)$ and $\tilde{c}_{ij}(k)$
are the 2D-Fourier transforms of the pair and direct correlation functions,
$h_{ij}(r)=g_{ij}(r)-1$ and $c_{ij}(r)$, where $g_{ij}(r)$ is the
radial distribution function between monomeric sites $i$ and $j$.
The 2D Fourier transform of a function $f(r)$ is defined as

\begin{equation}
\tilde{f}(k)=2\pi\int_{0}^{\infty}rdr\,f(r)J_{0}(kr)\label{ft2d}
\end{equation}
Since the Coulomb interaction is short ranged, the 2D Fourier transform
of the direct correlation functions are well defined at $k=0$, and
it is not necessary to take the special precautions described in Ref.\cite{13our2Dionic}
for unscreened Coulomb interactions.

The HNC closure equation is

\begin{equation}
g_{ij}(r)=\exp\left[-\beta v_{ij}(r)+h_{ij}(r)-c_{ij}(r)\right]\label{hnc}
\end{equation}
where the exponential term is missing the so-called bridge function
$b_{ij}(r)$ , which contains all the high order rank correlations.
Setting $b_{ij}(r)=0$ is the first level of controlled approximations,
which is why this particular closure is interesting. Other choices,
such as the mean spherical approximation (MSA)\cite{35hansmac}, represent
even higher level of approximation. Further choices, such as self-consistent
closures\cite{2995hansen,2997lomba}, or the Hirata-Kovalenko type
closure\cite{34KH}, are uncontrolled approximations, mostly based
in empirical methodologies, which could help improve finding numerical
solutions when HNC cannot, but such methods cannot help understand
nor appreciate the role played by high order correlation through $b_{ij}(r)$. 

Both the HNC and the SSOZ equations are approximations. In particular,
the SSOZ equation used here, has known deficiencies\cite{36Rism}.
It is only through the comparison with simulation that one can assert
the range of applicability of these 2 equations. This is the empirical
approach that we use here.

These 2 equations (\ref{ssoz}, \ref{hnc}) are iteratively solved
using standard techniques developed for the 2D case\cite{37Lado2D}.
The correlation functions are sampled on a logarithmic grid of 1024
points, and the Fourier transforms are handled through the Talman
technique\cite{40talman,41my2Delli}.

The atom-atom structure factors shown in Section 3 are defined as

\begin{equation}
S_{ij}(k)=1+\frac{\rho}{2}\tilde{h}_{ij}(k)\label{Sij}
\end{equation}
They are related to the structure factor defined in Eq(\ref{sk})
by removing the intramolecular part. Also we add 1 to the cross terms,
instead of the usual $\delta_{ij}$, in order to facilitate the graphical
representation.

While the simulations meet no problems even when strong clustering
is present, we find that the IET cannot be solved below the no-solution
line shown in Fig.2. As mentioned in Ref.\cite{13our2Dionic}, in
the case of Coulomb interactions, this behaviour does not appear to
be due to the onset of a liquid-gas transition, but to a strong clustering
of opposite charges. This clustering increases the first peak of the
unlike ions correlations, which is one of the causes for the raise
of the corresponding structure factor near k=0, in addition to the
usual long range tail of the correlations. These points are discussed
in the Results section below.

\section{Results}

\subsection{Phase diagram}

Fig.2 shows the no-solution ``phase'' diagram as obtained by the
HNC approximation. The data from the monomer ionic fluid of Ref.\cite{13our2Dionic}
is equally shown in dotted lines. It is seen that the, to the difference
of a factor in density, which could match the difference in volume
of the two types of system, the shape is nearly the same. The dimer
model shows an increase at high density which indicates could correspond
to the existence of the solid phase at even higher density. What is
more intriguing is the flat asymptote behaviour at very low densities,
as shown in the inset with densities in log scale. This very reminiscent
of the behaviour of the true Coulomb 2D system, which undergoes a
Kosterlitz-Thouless transition in the low density region, where pairs
of opposite charges bind and the system becomes electrically neutral
for lower temperatures\cite{2990fisher,2991pana,2995hansen}. In Ref.\cite{13our2Dionic},
our results suggested that the theory predicts something very similar,
as previously noted\cite{2997lomba}. The low density behaviour of
the present system is even more striking with the KT-like behaviour.
However, snapshots of simulations indicate that, in the vicinity of
the no-solution line, despite strong clustering, there are several
free charges, which can be explained by the fact that the interaction
is screened, and particles far away cannot irreversibly form pairs.
At lower temperatures, we observe that all charges form bilayer type
clusters and no free charges remain. So, it is tempting to conclude
that the approximate theory predicts this binding reminiscent of the
KT ion pairing.

The working hypothesis of this report is that the no-solution line
delimited by HNC corresponds to a physical line below which the simulations
show that clustering occurs. The simulation themselves do not show
any sort of singularity except for the existence of marked clustering
below this line. This is very similar to what we reported in Ref.\cite{13our2Dionic}.
At present we have no specific characterization of what the ``phase''
below the no-solution line could represent, other than ``cluster
phase''. As can be seen in the energy plot (Fig.4 discussed below),
there is no sign of thermodynamic singularity in the vicinity of the
line.

\subsection{Snapshots}

Fig.3a-d show typical snapshots of the system showing how clusters
form below the no-solution line, for 3 different densities ranging
from dense liquid $\rho=0.7$ to gas phase $\rho=0.01$. For each
density, 3 temperatures are shown, a high temperature (above and close
to the no-solution region), a temperature just below the no-solution
region, and the lowest temperature T=500K we simulated. At this temperature
we would expect a solid phase in principle. The contrast between the
disordered behaviour above the no-solution line and the existence
below it of well defined clusters with no or little free particles
is obvious and striking. For the dense liquid, we observe that below
the no-solution line charge and invert groups show micro-segregation.
In 3D it is possible to find solutions with a segregated domain pre-peak
in the structure factor, but not in 2D. 

The existence of ``droplets'' for lower densities could suggest
that the system is in a 2 phase region, implying a phase separation.
However, we could not observe such phase coexistence between a presumed
gas and a liquid. The fact that small pieces of bilayers are formed
in the dense region indicates that no liquid phase is formed. It is
more tempting to suggest the existence of a cluster phase rather.

\subsection{Thermodynamics}

Fig.4 shows the energies (upper panel) and the constant volume heat
capacities (lower panel) versus temperature. Each curve corresponds
to a isochore. In relation to the no-solution diagram in Fig.2, we
have drawn the curves for densities below $\rho=0.1$ in dotted lines,
and full lines for densities above. The curve for $\rho=0.4$ which
corresponds to the minimum of the no-solution line is drawn in red.
The two highest densities,$\rho=0.76$ and $\rho=0.7$ are shown in
dotted cyan, since they have trends different from the other curves.
The dots represent the no-solution line in Fig.2. For each density,
the energy is seen to become more negative as more and more clusters
form, which is expected. The heat capacity is seen to become more
noisy in the low temperature cluster region, which is expected of
the rather small size (N=200) in the simulations. In a way, the appearance
of these fluctuations help delimitate the cluster region. However,
we do not see any signature or singularity in the vicinity of the
no-solution line. 

It is easy to connect the no-solution points in the energy/temperature
diagram into a u-shaped curve, but not in the heat capacity diagram,
specially for the high density part.

The conclusion we draw is that usual thermodynamic quantities which
help signal first or second order transitions, do not show any singularities
or marked behaviour as the cluster line, or the no-solution of the
IET, are crossed.

\subsection{Correlation functions and structure factors}

Fig.5a-e show a comparison between the simulation and the HNC approximation
for all the 6 site-site correlation functions (left panel) and corresponding
structure factors (right panel), for different densities and temperatures
very close above the no-solution line of Fig.2, which are the most
demanding conditions to test the theory. These plots principally illustrate
the remarkable agreement between the simulation and IET data. For
the dense liquid phase $\rho=0.7$, we also show a plot for a high
temperature (Fig.5a), demonstrating that there is not much quantitative
difference with the low temperature case in Fig.5b. Fig.5e shows a
comparison for very low density $\rho=0.01$ at $T=3500$K, and more
particularly an interesting sub-structure which appears for $g_{+-}(r)$
obtained by the IET. It concerns a marked double shoulder feature
which appear just at base of the first peak around $r\approx2\sigma$.
This feature is absent from simulation (magenta curve). However, we
find a similar feature in the simulation data, but for a lower density
$\rho=0.002$. This is reported as a green curve in Fig.5e. This feature
gives an indirect indication in the clustering differences between
theory and simulations. We believe that the HNC closure tends to exaggerate
near neighbour correlations, as can be noted for the hard sphere fluid\cite{35hansmac},
but also for orientational ordering\cite{41my2Delli}. Following this
feature of the theory, and the slanting of the no-solution curve in
the low density region of Fig.2, it seems reasonable to suppose that
the clustering feature found in the theory happens at an lower density
for the same temperature. This finding points towards again towards
clustering dominating the lower part of the phase diagram. For this
low density $\rho=0.01$ in the left panel of Fig.5e, we has shown
the correlation functions $\tilde{h}_{ij}(k)$ instead of the $S_{ij}(k)$,
since the low density $\rho=0.01$ damps all the features because
of the definition Eq.(\ref{Sij}). All these figures show minor disagreements
here and there, but these can easily be accounted for the missing
bridge term in Eq.(\ref{hnc})

The figures also show that the very good agreement between the 2 approaches,
holds both in real and reciprocal space. Such agreement was equally
noticed in 3D in Ref.\cite{20ourIL}. It was attributed to low fluctuations
of the short range order and the subsequent homogeneity enforced by
the strong charge ordering. Indeed, approximate IET tend to be less
accurate when fluctuations are present, either as reflecting the great
possibilities of positional order, such as in simple liquids, or long
range correlations such as in the vicinity of phase transitions. In
contrast, the strong local order imposed by the charge ordering reduces
fluctuations and enforce the type of agreement we observe for all
the region where solutions could be found. The fact that the worse
agreement is precisely found for the correlations involving the neutral
site X, which have more disorder in their positioning, further enforces
the homogeneity argument presented here.

The fact that the agreement holds all the way until the no-solution
line is hit, is a very strong indication about the nature of the state
below the no-solution line. It suggests that this state is not due
to some mechanical instability of the upper homogeneous phase. If
it was, then we would see a progressive loss of agreement as we near
the no-solution line from above. On the contrary, this second phase
is due to clustering and not fluctuations, as illustrated by the snapshots
in Fig.2. The passage from homogeneous phase to cluster phase is not
made through any thermodynamic signature, such as heat capacity or
entropy, nor appearance of critical fluctuations. 

The approximate IET cannot account for the clusters which appear below
the no-solution line, because they miss high order correlation through
the so-called bridge function. However, the rather good agreement
with simulations found until the no-solution line, tend to indicate
that these bridge diagrams do not play an important role until this
line is met from above. It is possible that they become suddenly important
below this line, hence explaining why IET cannot get there. This tentative
explanation links the cluster ``phase'' to the raise in importance
of high rank correlations, principally through the bridge diagram
term. This points requires separate investigations.

\subsection{Supra-molecular structures and pre-peak}

In a previous study of 3D room temperature model ionic liquid by one
of us\cite{20ourIL}, it was found that the presence of neutral sites
induced a local segregation of charged and neutral sites, and in agreement
with what is observed in realistic such liquids\cite{24ppcastner}.
This segregation reflects itself in the presence of a low-k pre-peak
feature, observed both in the cross charge structure factor $S_{+-}(k)$
and in like charge structure factors $S_{++}(k)$ and $S_{--}(k)$.
By separating out the charge-charge and density-density structure
factors through the Bhatia-Thornton transformation\cite{50bt}, it
was found that only the density-density structure factor retained
this pre-peak\cite{20ourIL}, indicating that it is indeed related
to hetereogeneity in the spatial density distribution.

Fig.6 shows the Bhatia-Thornton (BT) structure factors of selected
state points lying just above the the no-solution line, and corresponding
to some of the structure factors shown in Fig.5b-e. These BT structure
factors are in fact related to a linear transformation from microscopic
densities of charged atom to total density and charge density. The
resulting structure factors are defined as in our previous work \cite{20ourIL}:

\begin{equation}
S_{NN}(k)=S_{++}(k)+S_{--}(k)+2S_{+-}(k)\label{BT}
\end{equation}

\[
S_{ZZ}(k)=\frac{1}{4}\left[z_{+}^{2}S_{++}(k)+z_{-}^{2}S_{--}(k)+2z_{+}z_{-}S_{+-}(k)\right]
\]
$S_{NN}(k)$ represent the structure factor related to total density
fluctuations, while charge fluctuations are represented by $S_{ZZ}(k)$.
We observe again that the agreement between the simulation and IET
data is very good. Similarly to Fig.5e, and again because of the low
density $\rho=0.01$, the lower right panel shows the $\tilde{h}(k)$
corresponding to the BT structure factor, with $\tilde{h}_{cc}(k)$
shifted by 1 in order to enforce the resemblance with the other plots.

These plots indicate that only the case $\rho=0.4$ shows marked pre-peak
feature in $S_{cc}(k)$, whereas the case for $\rho=0.7$ shows only
a shoulder, and the low density cases show mostly k=0 density fluctuations.
These finding are consistent with the snapshots shown in Fig.3a-d.
At high density, we observe a segregation of charged and neutral groups,
but the dimensionality does not allow for a marked segregation as
in the 3D case. The marked pre-peak for the medium density case is
possibly due to the clear clustering (Fig.3b), which enforces the
local heterogeneity. In Fig.3b, for T=1000K we observe clear chain-like
clusters with evident +- chain formation. In otherword, we see that
domain segregation is strongly affected by dimensionality, and that
it is stronger in 3D than in 2D. This finding could be of relevance
for the observation of charge segregation in adsorbed realistic RTILs.

\section{Discussion and Conclusion}

Although we use a theoretical approach, with simulations and integral
equations, this paper is similar to an experimental paper, in the
sense that we present only outcome of calculations, which in turn
suggest some properties of the system. In particular, we point to
the existence of a cluster phase, which is not accompanied by any
of the usual signatures for thermodynamical phase transitions. Since
we compare approximate theory with simulations, we cannot attest that
the no-solution line represents the actual boundary between the homogeneous
and the cluster phase. If it was possible to detect this line from
simulations alone, the evidence presented here suggests that it could
possibly lie below the approximate no-solution line predicted by IET,
but close to it. 

Although our results concern only screened version of the 3D Coulomb
interaction, we do not think that incorporating true 2D Coulomb interaction
would modify the conclusions reached here. This is because it is the
strong short range order is similar in both type of interactions,
and we conjecture that it rules the structural properties of this
type of systems.

Clustering plays an important role, even in simple liquids\cite{45wolfframCluster}.
It is usually relates to fluctuations, which concern principally the
$k=0$ part of the structure factor, as far as the stability of the
system is considered\cite{35hansmac}. Our experience in studying
realistic 3D associating liquids, such as water or alcohols, indicates
that fluctuations at $k\neq0$, in addition to being related to clustering\cite{47waterMH},
play little or no role in the global stability of the system. On the
contrary, they enhance stable local heterogeneity\cite{47waterMH}.
Therefore, since in the present case, simulations indicate that bilayer-like
clustering appears in the lower part of the phase diagram, they support
the fact that this system is governed by charge ordering induced clustering
everywhere in the phase diagram, albeit to various degrees. In view
of the remarkable agreement found in the correlation functions obtained
from simulation and calculated from the theory, these findings help
supporting the hypothesis that the no-solution line found in approximate
IET could be a physical line distinguishing between different clustering
regimes. The lower part of the phase diagram in Fig.2 would the be
dominated by many body correlations, which cannot be captured by two-body
level description. This line of argument would also explain why such
IET are unable to provide solutions for well mixed but micro-heterogeneous
aqueous mixtures, which could equally require explicit many body correlation
description. Subsequent investigation along these lines are in progress.

Finally, the polar-nonpolar domain segregation is found to be diminished
by dimensional reduction. This is very apparent for dense surface
coverage, but the segregation seems to be restored when particle confinement
conditions are decreased by lowering the surface coverage density.
This finding could have some relevance to 2D adsorption of realistic
3D ionic liquids.

\section*{Acknowledgments}

The authors thank the partenariat Hubert Curien (PHC) from Campus
France for financial support under the bilateral PROTEUS PHC project
35120VG.

\newpage

\part*{Figure Captions}
\begin{lyxlist}{00.00.0000}
\item [{Fig.1}] Ionic liquid model, with 1 site anion (red) and dimer cation
with 1 charged (blue) and 1 neutral (magenta). All sites have same
diameter and same 1/r\textasciicircum{}12 dispersive repulsion (see
text).
\item [{Fig.2}] (Density, temperature) no-solution ``phase'' diagram
from the IET. Yellow dots correspond to the lowest temperature for
which IET could be solved, and the blue line is connecting them. The
dotted line represent the no-solution line for Model 2 from Ref.\cite{13our2Dionic}.
The inset shows the same diagram but on log scale for the density.
\item [{Fig.3a}] Snapshots for high density $\rho=0.7$ at 3 different
temperatures $T=2000$K, $T=1000$K and $T=500$K. The anion is red,
cation blue and attached neutral site is magenta.
\item [{Fig.3b}] Snapshots for medium density $\rho=0.4$ at 3 different
temperatures $T=1000$K, $T=800$K and $T=50$0K
\item [{Fig.3c}] Snapshots for low density $\rho=0.1$ at 3 different temperatures
$T=2500$K, $T=1500$K and $T=500$K
\item [{Fig.3d}] Snapshots for very low density $\rho=0.01$ at 3 different
temperatures $T=3500$K, $T=2500$K and $T=50$0K
\item [{Fig.4}] Energy (top) and heat capacity (bottom) as a function of
temperature (divided by 1000), as obtained from the simulations. Each
line correspond to a density in the range $\rho=0.76$, $0.7$0, $0.60$,
$0.50$, $0.40$, $0.30$, $0.20$, $0.10$, $0.05$, $0.02$, $0.01$,
$0.00$5 and $0.002$. Curves for very low densities below $0.1$
are in dotted lines, as well as those for very high densities above
$0.65$ (cyan). The line in red is for $\rho=0.4$, which corresponds
to the minimum of the no-solution line of the IET in Fig2. Points
corresponding to this no-solution line are indicated in orange dots. 
\item [{Fig.5a}] Correlation functions (left) and corresponding structure
factors (right) for high density $\rho=0.7$ and high temperature
$T=3000$K. X designates the neutral site of the cation in Fig.1.
\item [{Fig.5b}] Same as Fig.5a, but for $\rho=0.7$ and temperature $T=1500$K
closer to the no-solution line in Fig.2.
\item [{Fig.5c}] Same as Fig.5a, but for medium density $\rho=0.4$ and
temperature $T=1001$K
\item [{Fig.5d}] Same as Fig.5a, but for low density $\rho=0.1$ and temperature
$T=2000$K
\item [{Fig.5e}] Same as Fig.5a, but for very low density $\rho=0.01$
and temperature $T=3500$K. The green curve is explained in the text.Note,
that it is $\tilde{h}_{ij}(k)$ that are plotted in the right panel
(see text)
\item [{Fig.6}] Bathia-Thornton structure factors $S_{cc}(k)$ and $S_{zz}(k)$
for the state points corresponding to Figs.5b-e. $S_{cc}(k)$ is shown
in blue for IET and dotted green for simulations. $S_{zz}(k)$ is
shown in red for IET and dotted gray for simulations. The lower right
panel shows $\tilde{h}_{cc}$ and $\tilde{h}_{zz}$ (see text).
\end{lyxlist}
.

\newpage

.

\includegraphics[scale=0.3]{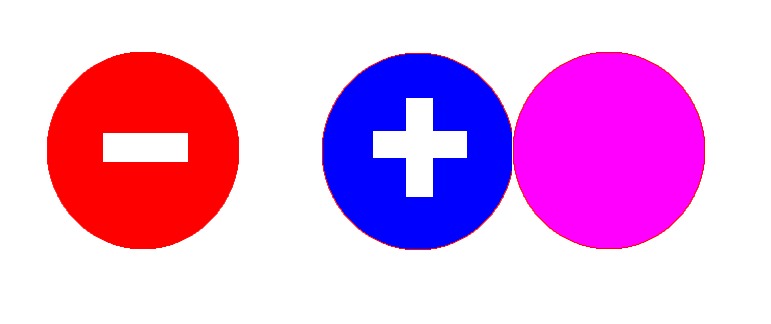}

.

Fig.1 - Ionic liquid model, with 1 site anion (red) and dimer cation
with 1 charged (blue) and 1 neutral (magenta). All sites have same
diameter and same 1/r\textasciicircum{}12 dispersive repulsion (see
text).

.

\newpage

.

\includegraphics[scale=0.7]{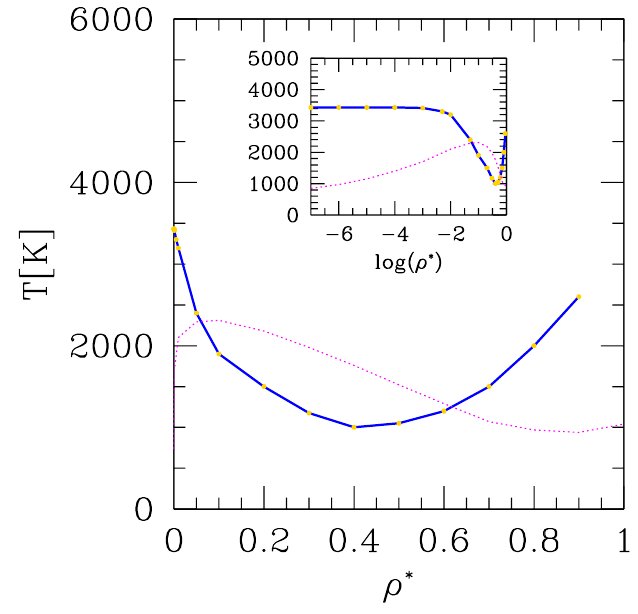}

.

. Fig.2 - (Density, temperature) no-solution ``phase'' diagram from
the IET. Yellow dots correspond to the lowest temperature for which
IET could be solved, and the blue line is connecting them. The dotted
line represent the no-solution line for Model 2 from Ref.\cite{13our2Dionic}.
The inset shows the same diagram but on log scale for the density.

\newpage

.

\includegraphics[scale=0.4]{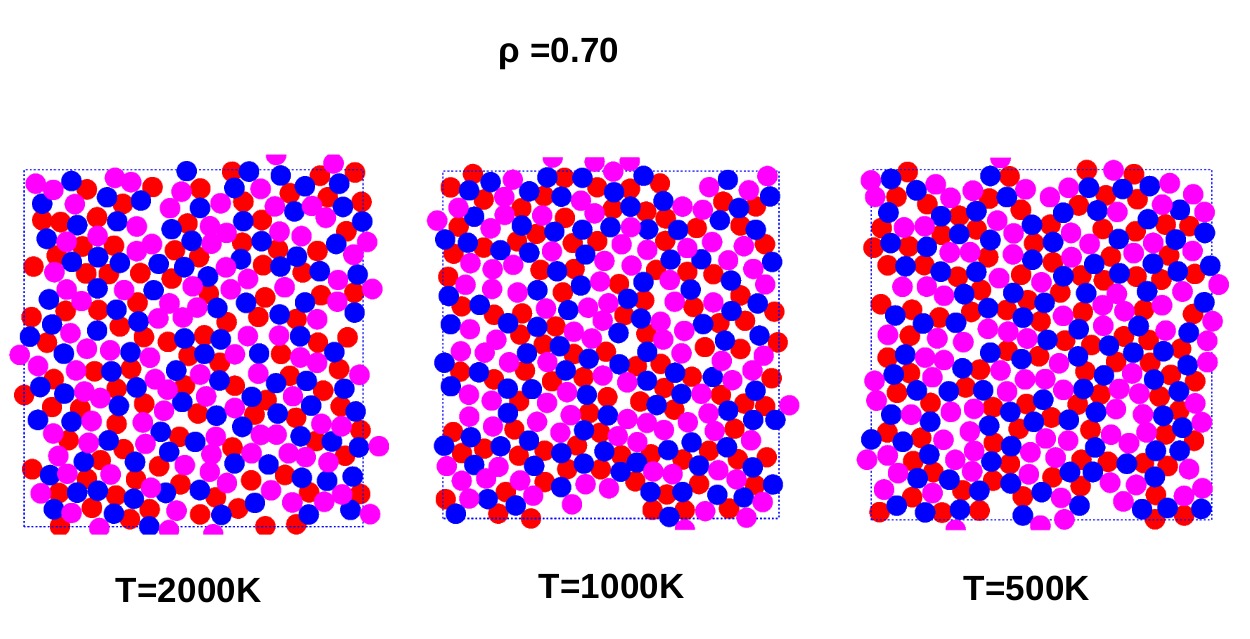}

.

. Fig.3a - Snapshots for high density $\rho=0.7$ at 3 different temperatures
$T=2000$K, $T=1000$K and $T=500$K. The anion is red, cation blue
and attached neutral site is magenta.

\newpage

.

\includegraphics[scale=0.4]{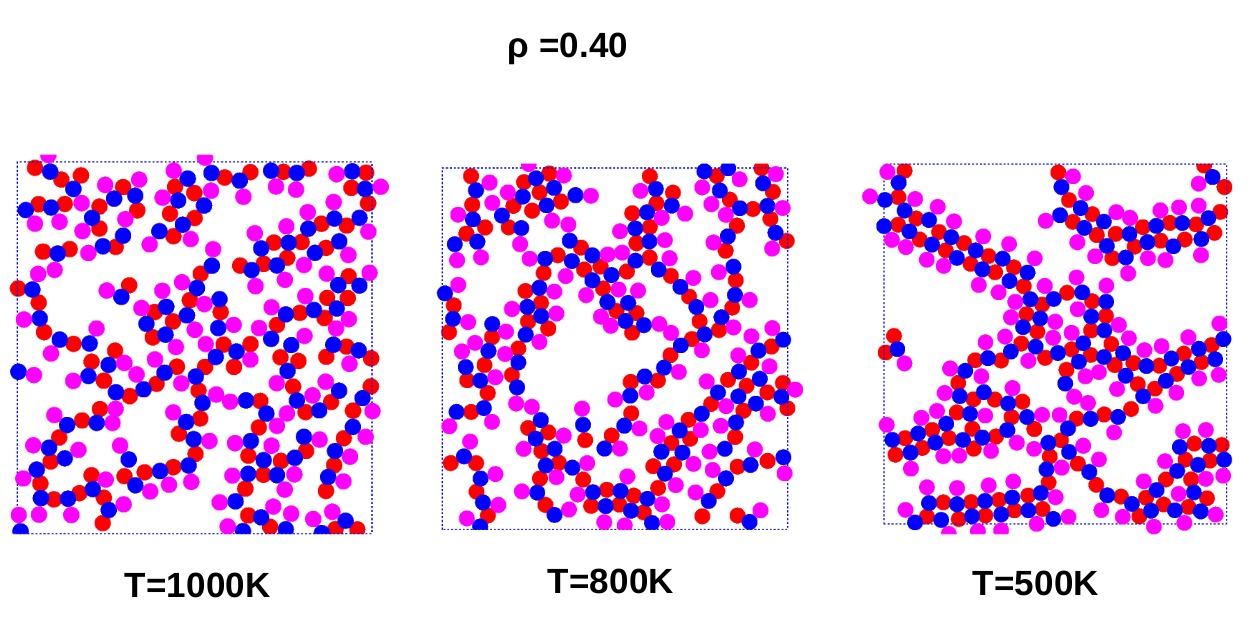}

.

.Fig.3b - Snapshots for medium density $\rho=0.4$ at 3 different
temperatures $T=1000$K, $T=800$K and $T=50$0K

\newpage

.

\includegraphics[scale=0.4]{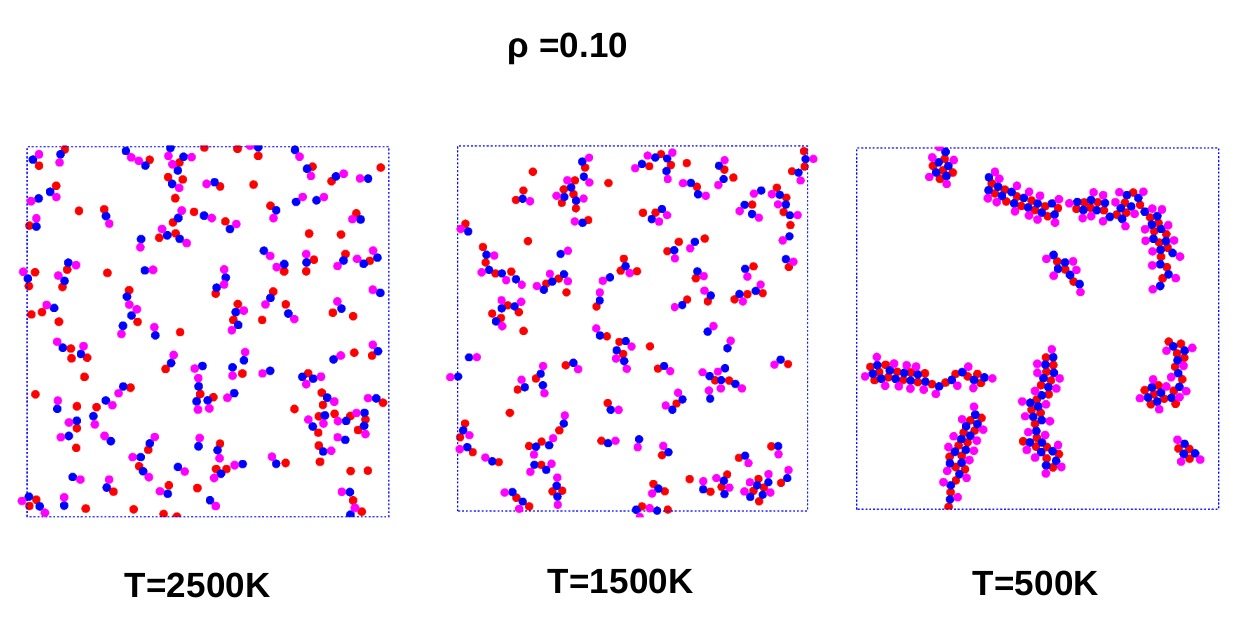}

.

.Fig.3c - Snapshots for low density $\rho=0.1$ at 3 different temperatures
$T=2500$K, $T=1500$K and $T=500$K

\newpage

.

\includegraphics[scale=0.4]{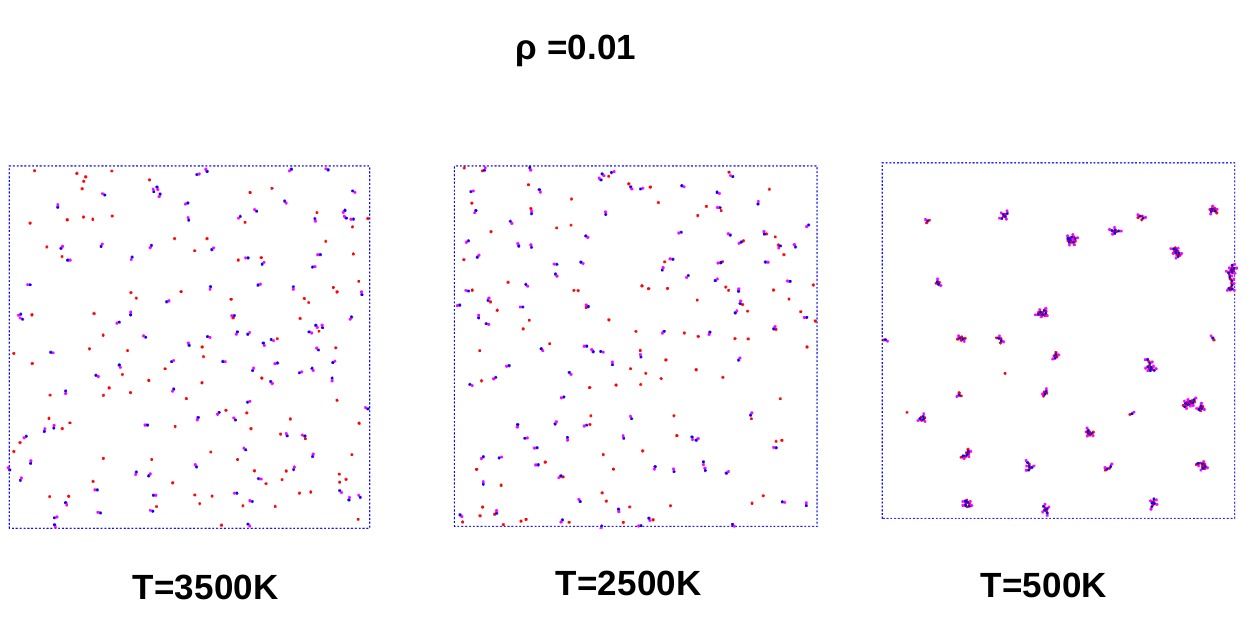}

.

.Fig.3d - Snapshots for very low density $\rho=0.01$ at 3 different
temperatures $T=3500$K, $T=2500$K and $T=50$0K

\newpage

.

\includegraphics[scale=0.6]{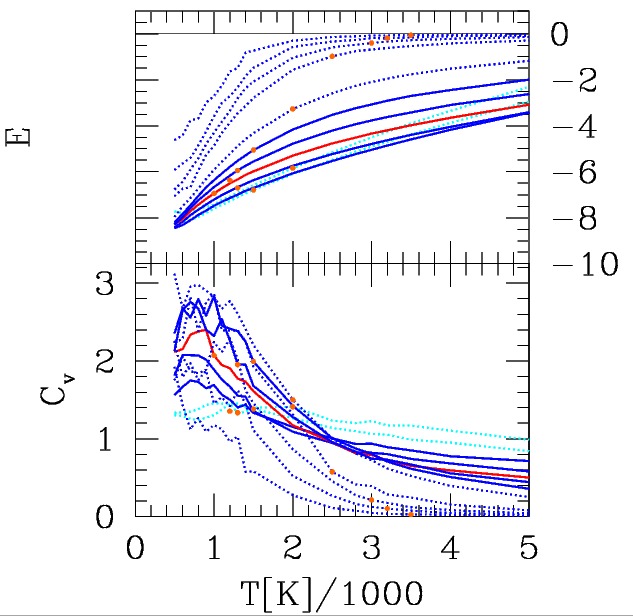}

.

.Fig.4 - Energy (top) and heat capacity (bottom) as a function of
temperature (divided by 1000), as obtained from the simulations. Each
line correspond to a density in the range $\rho=0.76$, $0.7$0, $0.60$,
$0.50$, $0.40$, $0.30$, $0.20$, $0.10$, $0.05$, $0.02$, $0.01$,
$0.00$5 and $0.002$. Curves for very low densities below $0.1$
are in dotted lines, as well as those for very high densities above
$0.65$ (cyan). The line in red is for $\rho=0.4$, which corresponds
to the minimum of the no-solution line of the IET in Fig2. Points
corresponding to this no-solution line are indicated in orange dots. 

\newpage

.

\includegraphics[scale=0.4]{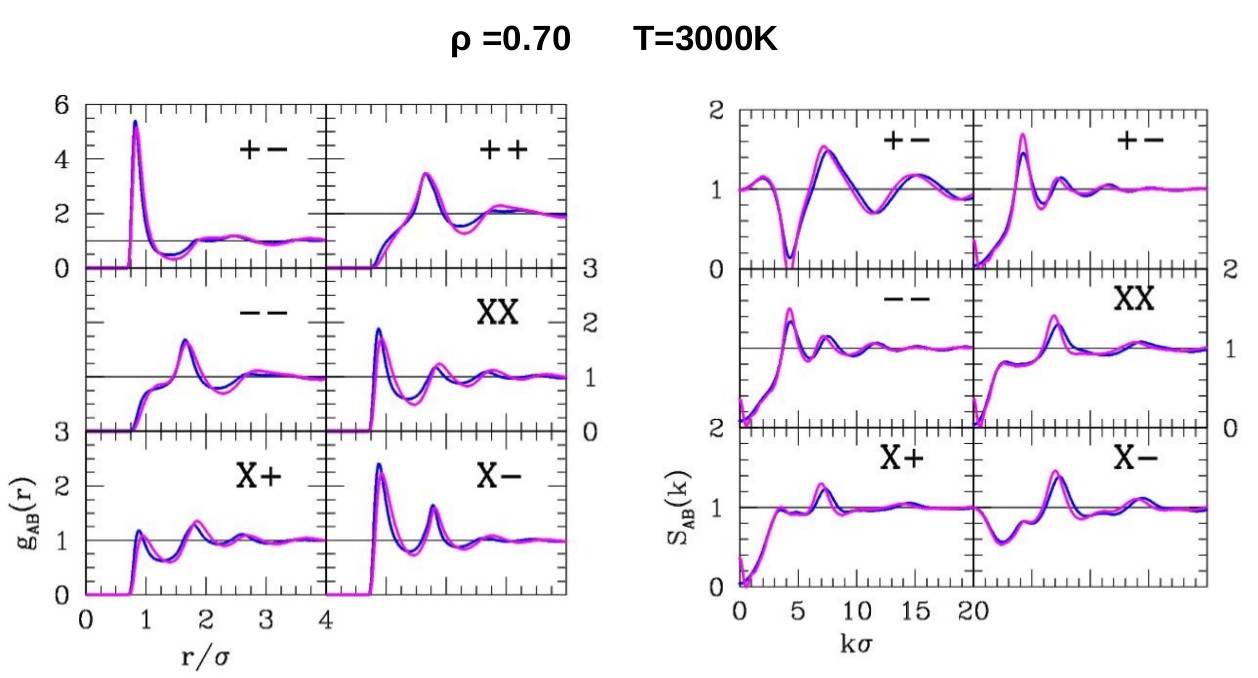}

.

.Fig.5a - Correlation functions (left) and corresponding structure
factors (right) for high density $\rho=0.7$ and high temperature
$T=3000$K. X designates the neutral site of the cation in Fig.1.

\newpage

.

\includegraphics[scale=0.4]{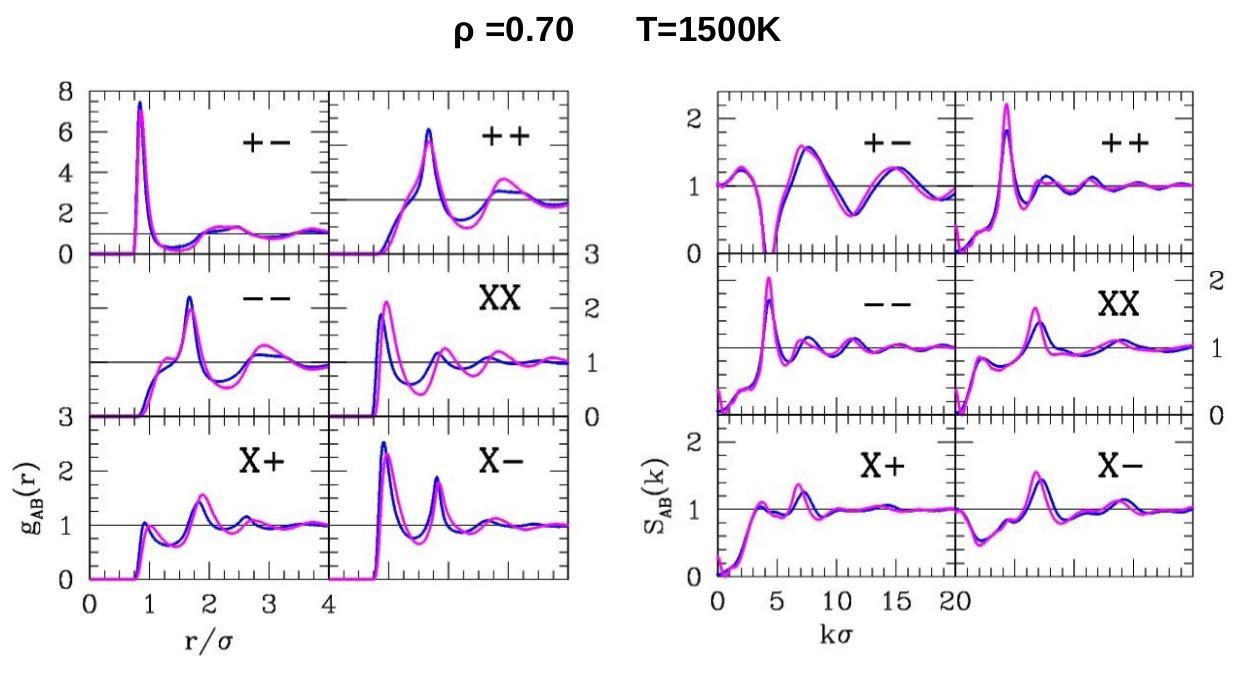}

.

.Fig.5b - Same as Fig.5a, but for $\rho=0.7$ and temperature $T=1500$K
closer to the no-solution line in Fig.2.

\newpage

.

\includegraphics[scale=0.4]{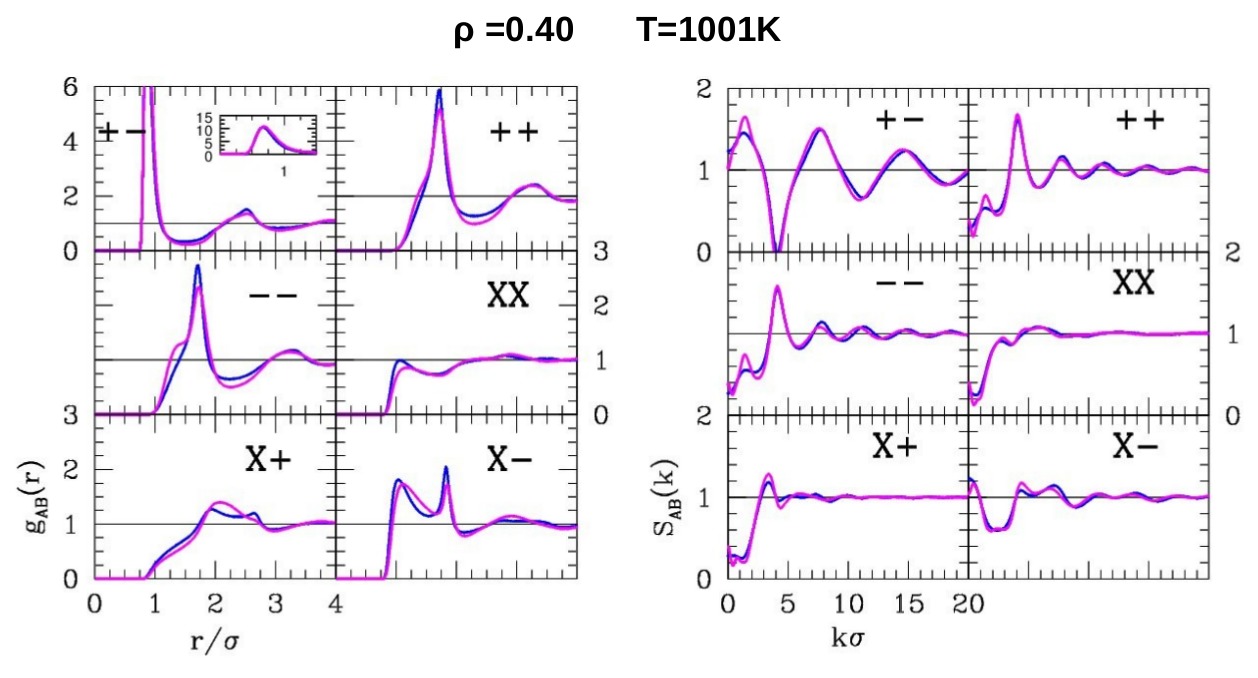}

.

.Fig.5c - Same as Fig.5a, but for medium density $\rho=0.4$ and temperature
$T=1001$K

\newpage

.

\includegraphics[scale=0.4]{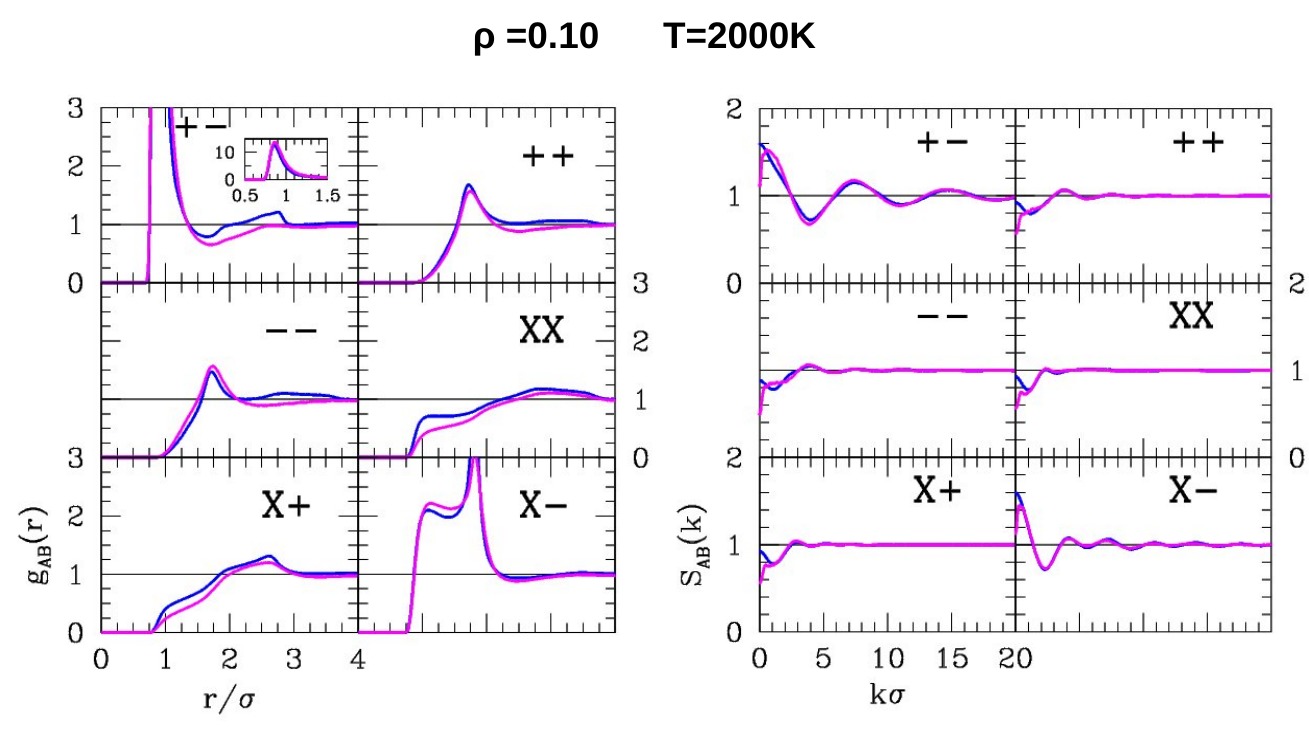}

.

.Fig.5d - Same as Fig.5a, but for low density $\rho=0.1$ and temperature
$T=2000$K

.

\newpage

.

\includegraphics[scale=0.4]{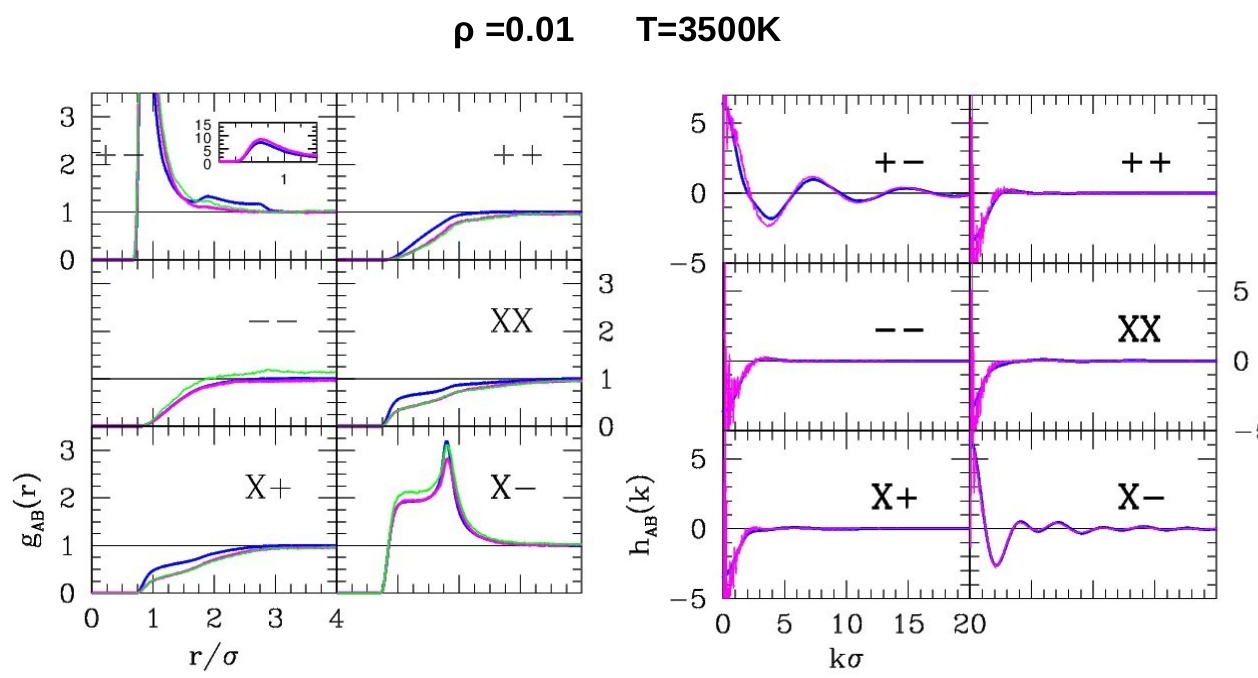}

.

.Fig.5e - Same as Fig.5a, but for very low density $\rho=0.01$ and
temperature $T=3500$K. The green curve is explained in the text.
Note, that it is $\tilde{h}_{ij}(k)$ that are plotted in the right
panel (see text)

.

\newpage

.

\includegraphics[scale=0.5]{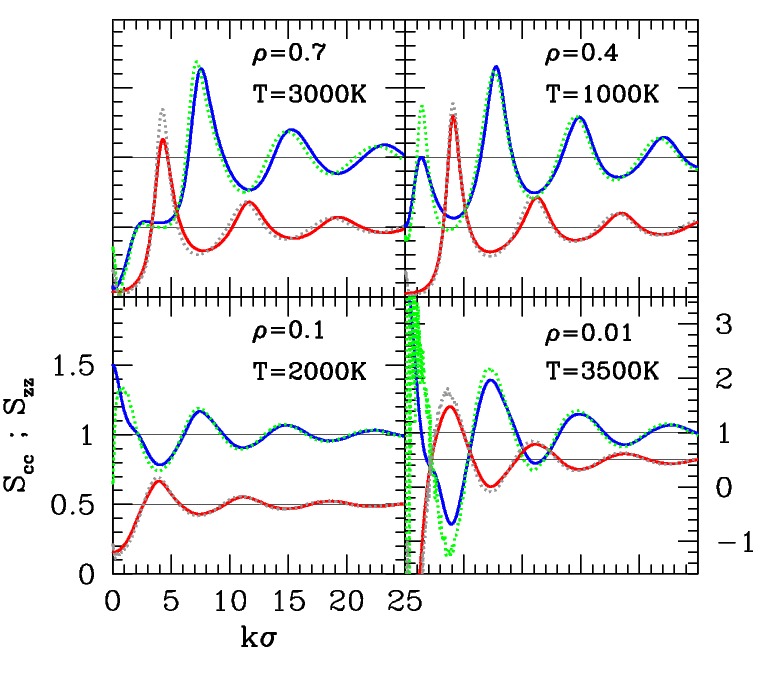}

.

.Fig.6 - Bathia-Thornton structure factors $S_{cc}(k)$ and $S_{zz}(k)$
for the state points corresponding to Figs.5b-e. $S_{cc}(k)$ is shown
in blue for IET and dotted green for simulations. $S_{zz}(k)$ is
shown in red for IET and dotted gray for simulations. The lower right
panel shows $\tilde{h}_{cc}$ and $\tilde{h}_{zz}$ (see text).
\end{document}